\documentclass[11pt]{article}

\usepackage{subfigure}
\usepackage{amsfonts}
\usepackage{amsmath}
\usepackage{amssymb}
\usepackage{graphicx}
\usepackage{fullpage}
\usepackage{algorithm}
\usepackage{algorithmic}

\newtheorem{theorem}{\textbf{Theorem}}
\newtheorem{lemma}{\textbf{Lemma}}

\newcommand{\eq}[1]{\begin{equation}\label{#1}}
\newcommand{\en}{\end{equation}}

\def\beq        {\begin{equation}}
\def\eeq        {\end{equation}}
\def\bea        {\begin{eqnarray}}
\def\eea        {\end{eqnarray}}

\title{Polynomial Filtering for Fast Convergence in Distributed Consensus}

\author{Effrosyni Kokiopoulou and Pascal Frossard\thanks{This work has been partly supported by the Swiss NSF, under NCCR
IM2.} \\
Ecole Polytechnique F\'ed\'erale de Lausanne (EPFL)\\
Signal Processing Laboratory (LTS4)\\
Lausanne - 1015, Switzerland\\
\tt{\{effrosyni.kokiopoulou,pascal.frossard\}@epfl.ch}}

\date{}

\begin{document}



\maketitle

\begin{abstract}

In the past few years, the problem of distributed consensus has
received a lot of attention, particularly in the framework of ad
hoc sensor networks. Most methods proposed in the literature
address the consensus averaging problem by distributed linear
iterative algorithms, with asymptotic convergence of the consensus
solution. The convergence rate of such distributed algorithms
typically depends on the network topology and the weights given to
the edges between neighboring sensors, as described by the network
matrix. In this paper, we propose to accelerate the convergence
rate for given network matrices by the use of polynomial filtering
algorithms. The main idea of the proposed methodology is to apply
a polynomial filter on the network matrix that will shape its
spectrum in order to increase the convergence rate. Such an
algorithm is equivalent to periodic updates in each of the sensors
by aggregating a few of its previous estimates. We formulate the
computation of the coefficients of the optimal polynomial as a
semi-definite program that can be efficiently and globally solved
for both static and dynamic network topologies. We finally provide
simulation results that demonstrate the effectiveness of the
proposed solutions in accelerating the convergence of distributed
consensus averaging problems.

\end{abstract}


\section{Introduction}

We consider the problem of distributed consensus
\cite{BertsekasTsitsiklis-book} that has become recently
very interesting especially in the context of ad hoc sensor networks. In particular, the problem of distributed average consensus has
attracted a lot of research efforts due to its numerous
applications in diverse areas. A few examples include
distributed estimation \cite{SchizRibGiannakis-TSP.08},
distributed compression \cite{RaHaSiNowak.06}, coordination of
networks of autonomous agents \cite{BloHendrOlshTsi.05} and
computation of averages and least-squares in a distributed fashion
(see e.g., \cite{XiaoBoyd.04,XiBoLa.05,XiBoLa.06,OlshTsi.06} and
references therein).

In general the main goal of distributed consensus is to reach a
global solution using only local computation and communication while staying robust to changes in the network topology.
Given the initial values at the sensors, the problem of
distributed averaging is to compute their average \textit{at each
sensor} using distributed linear iterations. Each distributed
iteration involves local communication among the sensors. In
particular, each sensor updates its own local estimate of the
average by a weighted linear combination of the corresponding
estimates of its neighbors. The weights that are represented in a network weight matrix $W$ typically drive the importance of
the measurements of the different neighbors.

One of the important characteristics of the distributed consensus algorithms is the rate of convergence to the asymptotic solution.
In many cases, the average consensus solution can be reached by successive multiplications of
$W$ with the vector of initial sensor values. Furthermore, it has been
shown in \cite{XiaoBoyd.04} that in the case of fixed network
topology, the convergence rate depends on the second largest
eigenvalue of $W$, $\lambda_2(W)$. In particular, the convergence is faster when the value of
$\lambda_2(W)$ is small. Similar convergence results have been proposed recently in the case of random network
topology \cite{KarMoura.07,KarMoura.TSP.07}, where the convergence rate is governed by the expected value of the $\lambda_2(W)$, $E[ \lambda_2(W)]$.

The main research direction so far focuses on the computation of
the optimal weights $W$ that yield the fastest convergence rate to
the consensus solution \cite{XiaoBoyd.04,XiBoLa.05,XiBoLa.06}. In
this work, we diverge from methods that are based on successive
multiplications of $W$, and we rather allow the sensors to use
their previous estimates, in order to accelerate the convergence
rate. This is similar in spirit to the works proposed
\cite{KokFro.07.SPL,SunHadji.07} that reach the consensus solution
in a finite number of steps. They use respectively extrapolation
methods and linear dynamical system formulation for fixed network
topologies. In order to address more generic network topologies,
we propose here to use a matrix polynomial $p$ applied on the
weight matrix $W$ in order to shape its spectrum. Given the fact
that the convergence rate is driven by $\lambda_2(W)$, it is
therefore possible to impact on the convergence rate by careful
design of the polynomial $p$. In the implementation viewpoint,
working with $p(W)$ is equivalent to each sensor aggregating its
value periodically using its own previous estimates. We further
formulate the problem of the computation of the optimal
coefficients for both static and dynamic network topologies. We
show that this problem can be efficiently and globally solved in
both cases by the definition of a semi-definite program (SDP).

The rest of this paper is organized as follows. In Section
\ref{sec:background} we review the main convergence results of
average consensus in both fixed and dynamic random network
topologies. Next, in Section \ref{sec:polfilt} we introduce the
polynomial filtering methodology and discuss its implementation for distributed consensus problems. We compute the optimal polynomial filter in Section \ref{sec:polopt} for both static and dynamic network topologies. In Section
\ref{sec:experiments} we provide simulation results that verify
the validity and the effectiveness of our method. Related work is finally presented in Section
\ref{sec:related}.

\section{Convergence in Distributed Consensus Averaging}
\label{sec:background}

Let us first define formally the problem of distributed consensus averaging. Assume that initially each sensor $i$
reports a scalar value $x_0(i) \in \mathbb{R}$. We denote by $x_0 = [x_0(1),\ldots,x_0(n)]^\top \in \mathbb{R}^n$ the vector of
initial values on the network. Denote by
\begin{eqnarray}\label{eq:avg}
\mu = \frac{1}{n} \sum_{i=1}^n x_0(i)
\end{eqnarray}
the average of the initial values of the sensors. However, one rarely has a complete view of the network. The problem of
distributed averaging therefore becomes typically to compute $\mu$ \textit{at each sensor}
by distributed linear iterations. In what follows we review the
main convergence results for distributed consensus algorithms on both fixed and random network
topologies.

\subsection{Static network topology}

We model the static network topology as an undirected graph
$\mathcal{G} = (\mathcal{V},\mathcal{E})$ with nodes $\mathcal{V}
= \{1,\ldots,n\}$ corresponding to sensors. An edge $(i,j) \in
\mathcal{E}$ is drawn if and only if sensor $i$ can communicate
with sensor $j$. We denote the set of neighbors for node $i$ as
$\mathcal{N}_i = \{j|~ (i,j) \in \mathcal{E}\}$. Unless otherwise
stated, we assume that each graph is simple i.e., no loops or
multiple edges are allowed.

In this work, we consider distributed linear iterations of the
following form
\begin{equation}\label{eq:distliniter1}
x_{t+1}(i) = W_{ii} x_{t}(i) + \sum_{j \in \mathcal{N}_i} W_{ij}
x_t(j),
\end{equation}
for $i = 1,\ldots,n$, where $x_t(j)$ represents the value computed
by sensor $j$ at iteration $t$. Since the sensors communicate in
each iteration $t$, we assume that they are synchronized. The
parameters $W_{ij}$ denote the edge weights of $\mathcal{G}$. Since each sensor communicates only with its direct neighbors,
$W_{ij} = 0$ when $(i,j) \notin \mathcal{E}$. The above
iteration can be compactly written in the following form
\begin{equation}\label{eq:distliniter2}
x_{t+1} = W x_t ,
\end{equation}
or more generally
\begin{equation}\label{eq:x_t}
x_t = \Big( \prod_{i=0}^{t-1} W \Big) x_0 = W^{t} x_0.
\end{equation}

We call the matrix $W$ that gathers the edge weights $W_{ij}$, as
the weight matrix. Note that $W$ is a sparse matrix whose sparsity
pattern is driven by the network topology. We assume that $W$ is
symmetric, and we denote its eigenvalue decomposition as $W = Q \Lambda Q^\top$. The (real) eigenvalues can further be arranged as follows:
\begin{eqnarray}
\lambda_1(W) \geq \lambda_2(W) \geq \ldots \geq \lambda_{n-1}(W)
\geq \lambda_n(W).
\end{eqnarray}

The distributed linear iteration given in eq. (\ref{eq:distliniter2})
converges to the average if and only if
\begin{eqnarray}\label{eq:condW}
\lim_{t \rightarrow \infty} W^t = \frac{\textbf{1}
\textbf{1}^\top}{n},
\end{eqnarray}
where $\textbf{1}$ is the vector of ones \cite{XiaoBoyd.04}. Indeed, notice that in
this case \[x^\ast = \lim_{t \rightarrow \infty} x_t = \lim_{t
\rightarrow \infty} W^t x_0 = \frac{\textbf{1} \textbf{1}^\top}{n}
x_0 = \mu \textbf{1}.\]

It has been shown that for fixed network topology the convergence rate
of eq. (\ref{eq:distliniter2}) depends on the magnitude of the second
largest eigenvalue $\lambda_2(W)$ \cite{XiaoBoyd.04}. The asymptotic convergence factor is defined as
\begin{eqnarray}\label{eq:asymptotic}
r_{\textrm{asym}}(W) = \sup_{x_0 \neq \mu \textbf{1}} \lim_{t
\rightarrow \infty} \Big( \frac{\| x_t - \mu \textbf{1} \|_2}{ \|
x_0 - \mu \textbf{1}  \|_2} \Big)^{1/t},
\end{eqnarray}
and the per-step convergence factor is written as
\begin{eqnarray}\label{eq:perstep}
r_{\textrm{step}}(W) = \sup_{x_0 \neq \mu \textbf{1}} \frac{\|
x_{t+1} - \mu \textbf{1} \|_2}{ \| x_t - \mu \textbf{1}  \|_2}.
\end{eqnarray}

Furthermore, it has been shown that the convergence rate relates to the spectrum of $W$, as given by the following theorem \cite{XiaoBoyd.04}.
\begin{theorem}\label{thm:convfixed}
The convergence given by eq. (\ref{eq:condW}) is guaranteed if and only if
\begin{eqnarray}
\textbf{1}^\top W &=& \textbf{1}^\top, \\
     W \textbf{1} &=& \textbf{1}, \\
\rho \big(W - \frac{\textbf{1} \textbf{1}^\top}{n} \big) &<& 1,
\end{eqnarray}
where $\rho(\cdot)$ denotes the spectral radius of a matrix.
Furthermore,
\begin{eqnarray}
r_{\textrm{asym}}(W) = \rho \big(W - \frac{\textbf{1} \textbf{1}^\top}{n}\big), \label{eq:condasym} \\
r_{\textrm{step}}(W) = \| W - \frac{\textbf{1} \textbf{1}^\top}{n}
\label{eq:condstep} \|_2.
\end{eqnarray}
\end{theorem}
According to the above theorem, $\frac{1}{\sqrt{n}}\textbf{1}$ is
a left and right eigenvector of $W$ associated with the eigenvalue
one, and the magnitude of all other eigenvalues is strictly less
than one. Note finally, that since $W$ is symmetric,
the asymptotic convergence factor coincides with the per-step
convergence factor, which implies that the relations (\ref{eq:condasym}) and
(\ref{eq:condstep}) are equivalent.

We give now an alternate proof of the above theorem that illustrates the importance of the second largest eigenvalue in the convergence rate. We expand the
initial state vector $x_0$ to the orthogonal eigenbasis $Q$ of
$W$; that is,
\[
x_0 =  \nu_1 \frac{\textbf{1}}{\sqrt{n}} + \sum_{j=2}^n  \nu_j
q_j,
\]
where $\nu_1 = \langle \frac{\textbf{1}}{\sqrt{n}}, x_0 \rangle$
and $\nu_j = \langle q_j, x_0 \rangle$. We further assume that
$\nu_1 \neq 0$. Then, eq. (\ref{eq:x_t}) implies that
\begin{eqnarray}
x_t &=& W^t x_0 = W^t \big(\frac{\nu_1}{\sqrt{n}} \textbf{1} + \sum_{j=2}^n \nu_j q_j \big) \nonumber \\
    &=& \frac{\nu_1}{\sqrt{n}} \textbf{1} +  \sum_{j=2}^n \nu_j W^t q_j \nonumber \\
    &=& \frac{\nu_1}{\sqrt{n}} \textbf{1} +  \sum_{j=2}^n \nu_j \lambda_j^t q_j.  \nonumber
\end{eqnarray}
Observe now that if $|\lambda_j| < 1, ~\forall j \geq 2$, then in
the limit, the second term in the above equation decays and
\[
x^\ast = \lim_{t \rightarrow \infty} x_t = \frac{\nu_1}{\sqrt{n}}
\textbf{1} = \mu \textbf{1}.
\]

We see that the smaller the value of $\lambda_2(W)$, the faster the convergence rate. Analogous convergence results hold in the case of dynamic network
topologies discussed next.

\subsection{Dynamic network topology}\label{sec:randomtopology}

Let us consider now networks with random link failures, where the
state of a link changes over the iterations. In particular, we use
the random network model proposed in
\cite{KarMoura.07,KarMoura.TSP.07}. We assume that the network at
any arbitrary iteration $t$ is $\mathcal{G}(t) =
(\mathcal{V},\mathcal{E}(t))$, where $\mathcal{E}(t)$ denotes the
edge set at iteration $t$, or equivalently at time instant $t$. Since the network is dynamic, the edge
set changes over the iterations, as links fail at random. We
assume that $\mathcal{E}(t) \subseteq \mathcal{E}^\ast$, where
$\mathcal{E}^\ast \subseteq \mathcal{V} \times \mathcal{V}$ is the
set of realizable edges when there is no link failure.

We also assume that each link $(i,j)$ fails with a probability $(1-p_{ij})$,  independently of the
other links. Two random edge sets $\mathcal{E}(t_1)$ and $\mathcal{E}(t_2)$ at different iterations $t_1$ and $t_2$ are
independent. The probability of forming a particular $\mathcal{E}(t)$
is thus given by $\prod_{(i,j) \in \mathcal{E}(t)} p_{ij}$. We define the
matrix $P$ as
\begin{equation}
P_{ij} =
\begin{cases}
p_{ij}  & \textrm{if} ~(i,j) \in \mathcal{E}^\ast ~\textrm{and}~ i \neq j, \\
0                   & \textrm{otherwise}.
\end{cases}
\end{equation}
The matrix $P$ is symmetric and its diagonal elements are zero,
since it corresponds to a simple graph. It represents the probabilities of edge formation in the network, and the edge set
$\mathcal{E}(t)$ is therefore a random subset of $\mathcal{E}^\ast$ driven by the  $P$ matrix. Finally, the weight matrix $W$ becomes dependent on the edge set since only the  weights of existing edges can take non zero values.

In the dynamic case, the distributed linear iteration of eq.
(\ref{eq:distliniter1}) becomes
\begin{equation}\label{eq:distliniter1dyn}
x_{t+1}(i) = W_{ii}(t) x_{t}(i) + \sum_{j \in \mathcal{N}_i}
W_{ij}(t) x_t(j)
\end{equation}
or in compact form,
\begin{equation}\label{eq:distliniter2dyn}
x_{t+1} = W_t x_t,
\end{equation}
where $W_t$ denotes the weight matrix corresponding to the graph
realization $\mathcal{G}(t)$ of iteration $t$. The iterative relation given by eq. (\ref{eq:distliniter2dyn}) can be written as
\[
x_t = \Big( \prod_{i=0}^{t-1} W_i \Big) x_0.
\]
Clearly, $x_t$ now represents a stochastic process since the edges are drawn randomly.
The convergence rate to the consensus solution therefore depends on the behavior of the product $\prod_{i=0}^{t-1} W_i$.
We say that the algorithm converges if
\begin{equation}\label{eq:mss}
\forall x_0 \in \mathbb{R}^{n},~ \lim_{t \rightarrow \infty} E
\|x_t - \mu \textbf{1}\| = 0.
\end{equation}

We review now some convergence results from \cite{KarMoura.TSP.07}, which first shows that
\begin{lemma}
For any $x_0 \in  \mathbb{R}^{n}$,
\[
\|x_{t+1} - \mu \textbf{1}\| \leq \Big( \prod_{i=0}^t \rho \big(
W_i- \frac{\textbf{1}\textbf{1}^\top}{n} \big) \Big) \|x_{0} - \mu
\textbf{1}\|.
\]
\end{lemma}
It leads to the following convergence theorem \cite{KarMoura.TSP.07} for dynamic networks.
\begin{theorem}\label{thm:convdyn}
If $E[\rho\big( W - \frac{\textbf{1}\textbf{1}^\top}{n} \big)
\Big)] < 1$, the vector sequence $\{x_t\}_{t=0}^\infty$ converges
in the sense of eq. (\ref{eq:mss}).
\end{theorem}

We define the convergence factor in dynamic network topologies as
\[
r_d(W) = E \Big[ \rho \big(W - \frac{\textbf{1}
\textbf{1}^\top}{n} \big) \Big] . \label{eq:mssconvfactor}
\]
This factor depends
in general on the spectral properties of the induced network
matrix and drives the convergence rate of eq. (\ref{eq:distliniter2dyn}). The authors in \cite{TahJadba.06}
show that $|\lambda_2(E[W])| < 1$ is also a necessary and sufficient
condition for asymptotic (almost sure) convergence of the
consensus algorithm in the case of random networks, where both
network topology and weights are random (in particular i.i.d and
independent over time).

Finally, it is interesting to note that the consensus problem in a random network relates to
gossip algorithms. Distributed averaging under the synchronous gossip constraint implies that multiple node pairs may communicate
simultaneously only if these node pairs are disjoint. In other words, the set of
links implied by the active node pairs forms a matching of the
graph. Therefore, the distributed averaging problem described above is closely related to
the distributed synchronous algorithm under the gossip constraint
that has been proposed in \cite[Sec. 3.3.2]{BoydGhoshPrabShah.06}. It has been shown in this case that the averaging time (or
convergence rate) of a gossip algorithm depends on the second
largest eigenvalue of a doubly stochastic network matrix.


\section{Accelerated consensus with polynomial filtering}\label{sec:polfilt}

\subsection{Exploiting memory}

As we have seen above, the convergence rate of the distributed consensus algorithms depends
in general on the spectral properties of an induced network
matrix. This is the case for both fixed and random network
topologies. Most of the research work has been devoted to finding weight matrix $W$ for accelerating the convergence to the consensus solution when sensors only use their current estimates. We choose a different approach where we exploit the memory of sensors, or the values of previous estimates in order to augment to convergence rate.

Therefore, we have proposed in our previous work \cite{KokFro.07.SPL} the Scalar
Epsilon Algorithm (SEA) for accelerating the convergence rate to the consensus solution. SEA belongs to the family of extrapolation methods for
accelerating vector sequences, such as eq. (\ref{eq:distliniter2}).
These methods exploit the fact that the fixed point of the
sequence belongs to the subspace spanned by any $\ell+1$
consecutive terms of it, where $\ell$ is the degree of the minimal
polynomial of the sequence generator matrix (for more details, see
\cite{KokFro.07.SPL} and references therein). SEA is a low
complexity algorithm, which is ideal for sensor networks and it is
known to reach the consensus solution in $2\ell$ steps. However, $\ell$ is unknown in practice, so one may use all the
available terms of the vector sequence. Hence, the memory
requirements of SEA are $O(T)$, where $T$ is the number of terms.
Moreover, SEA assumes that the sequence generator matrix (e.g.,
$W$ in the case of eq. (\ref{eq:distliniter2})) is fixed, so that it does not adapt easily to dynamic network topologies.

In this paper, we propose a more flexible algorithm based on the
polynomial filtering technique. Polynomial filtering permits to
``shape" the spectrum of a certain symmetric weight matrix, in
order to accelerate the convergence to the consensus solution.
Similarly to SEA, it allows the sensors to use the value of their
previous estimates. However, the polynomial filtering methodology
introduced below presents three main advantages: (i) it is robust
to dynamic topologies (ii) it has explicit control on the
convergence rate and (iii) its memory requirements can be adjusted
to the memory constraints imposed by the sensor.

\subsection{Polynomial filtering}

Starting from a given (possibly optimal) weight matrix $W$, we
propose the application of a polynomial filter on the spectrum of
$W$ in order to impact the magnitude of $\lambda_2(W)$ that mainly drives the convergence rate. Denote by
$p_k(\lambda)$ the polynomial filter of degree $k$ that is applied
on the spectrum of $W$,
\begin{eqnarray}\label{eq:poly}
p_k(\lambda) = \alpha_0  + \alpha_1 \lambda + \alpha_2 \lambda^2 +
\ldots + \alpha_k \lambda^k.
\end{eqnarray}
Accordingly, the matrix polynomial is given as
\begin{eqnarray}\label{eq:polyW}
p_k(W) = \alpha_0 I  + \alpha_1 W + \alpha_2 W^2 + \ldots +
\alpha_k W^k.
\end{eqnarray}
Observe now that
\begin{eqnarray}
p_k(W) &=& p_k(Q \Lambda Q^\top) \nonumber \\
       &=& \alpha_0 I  + \alpha_1 (Q \Lambda Q^\top) + \ldots + \alpha_k (Q \Lambda^k Q^\top) \nonumber \\
       &=& Q p_k(\Lambda) Q^\top,\label{eq:pf}
\end{eqnarray}
which implies that the eigenvalues of $p_k(W)$ are simply the
polynomial filtered eigenvalues of $W$ i.e.,
$p_k(\lambda_i(W)),~i=1,\ldots,n$.

In the implementation level, working on $p_k(W)$ implies a
periodic update of the current sensor's value with a linear
combination of its previous values. To see why this is true,
we observe that:
\begin{eqnarray}
x_{t+k+1} &=& p_k(W) x_t \nonumber \\
          &=& \alpha_0 x_t  + \alpha_1 W x_t + \ldots + \alpha_k W^k x_t \nonumber \\
          &=& \alpha_0 x_t  + \alpha_1 x_{t+1} + \ldots + \alpha_k x_{t+k}
          \label{eq:pfupdate}.
\end{eqnarray}
A careful design of $p_k$ may impact the
convergence rate dramatically. Then, each sensor typically applies polynomial
filtering for distributed consensus by following the main steps
tabulated in Algorithm \ref{Algo:PFConsensus}.

\begin{algorithm}[tb]
\caption{Polynomial Filtered Distributed Consensus}
\label{Algo:PFConsensus}
\begin{algorithmic} [1]
\STATE \textbf{Input}: polynomial coefficients
$\alpha_0,\ldots,\alpha_{k+1}$, tolerance $\epsilon$. \STATE
\textbf{Output}: average estimate $\bar{\mu}$. \STATE
\textbf{Initialization}: \STATE \quad $\bar{\mu}_0 = x_0(i)$.
\STATE \quad Set the iteration index $t = 1$. \REPEAT
\IF{mod($t,k+1$) ==0}
\STATE $x_t(i) = \alpha_0 x_{t-k-1}(i)  + \alpha_1 x_{t-k}(i) +
\alpha_2 x_{t-k+1}(i) + \ldots + \alpha_k x_{t-1}(i)$
\COMMENT{polynomial filtered update}
\STATE $x_t(i) = W_{ii} x_{t}(i) + \sum_{j \in \mathcal{N}_i}
W_{ij}x_{t}(j)$. \ELSE
\STATE $x_t(i) = W_{ii} x_{t-1}(i) + \sum_{j \in \mathcal{N}_i}
W_{ij}x_{t-1}(j)$. \ENDIF \STATE Increase the iteration index $t =
t + 1$. \STATE $\bar{\mu}_t = x_t(i)$. \UNTIL{$\bar{\mu}_t -
\bar{\mu}_{t-1} < \epsilon$}
\end{algorithmic}
\end{algorithm}

Note that, for both fixed and random network topology cases, the $\alpha_k$'s
are computed off-line assuming that $W$ and respectively $E
[W]$ are known a priori. In what follows, we propose
different approaches for computing the coefficients $\alpha_i$ of the filter $p_k$.



\subsection{Newton's interpolating polynomial}

One simple and rather intuitive approach for the design of the polynomial
$p_k(\lambda)$ is to use Hermite interpolation. Recall that the objective is to dampen the
smallest eigenvalues $\lambda_2,\ldots, \lambda_n$ of $W$, while keeping the eigenvalue one intact. Therefore, we assume that the
spectrum of $W$ lies in an interval $[a,1]$ and we impose
smoothness constraints of $p_k$ at the left endpoint $a$. In
particular, the polynomial $p_k(\lambda): [a,1] \rightarrow
\mathbb{R}$ that we seek, will be determined by the following
constraints:
\begin{eqnarray}
p_k(a) &=& 0, ~ p_k^{(i)}(a) = 0, ~i = 1, \ldots, k-1 \label{eq:smoothness} \\
p_k(1) &=& 1, \label{eq:one}
\end{eqnarray}
where $p_k^{(i)}(a)$ denotes the $i$-th derivative of
$p_k(\lambda)$ evaluated at $a$. The dampening is achieved by imposing smoothness constraints of
the polynomial on the left endpoint of the interval. The computed polynomial will have
degree equal to $k$. Finally, the coefficients of $p_k$ that satisfies the
above constraints can be computed by the Newton's divided
differences table.

\begin{figure}[!t]
\begin{center}
\includegraphics[width=3in]{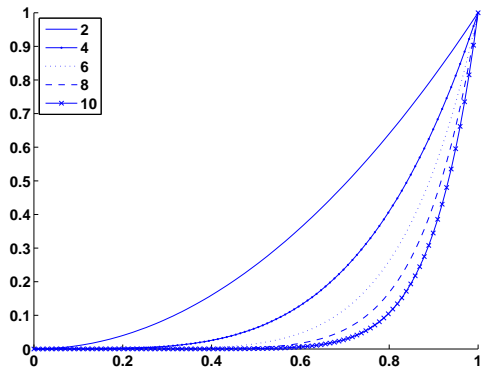}
\end{center}
\caption{Newton's polynomial of various degrees $k$.}
\label{fig:Newtonpol}
\end{figure}

Figure \ref{fig:Newtonpol} shows the form of
$p_k(\lambda)$ for $a = 0$ and different values of the degree $k$.
As $k$ increases, the dampening of the small eigenvalues becomes more
effective. It is worth mentioning that the design of Newton's polynomial does
not depend on the network size or the size of the weight matrix
$W$. What is only needed is a left endpoint $a$, which encloses
the spectrum of $W$, as well as the desired degree $k$, which
moreover may be imposed by memory constraints. This feature of
Newton's polynomial is very interesting and it is particularly
appealing in the case of dynamic network topologies. However, the above polynomial design is
mostly driven by intuitive arguments, which tend to obtain small
eigenvalues for faster convergence. In the following section, we
provide an alternative technique for computing the polynomial
filter that optimizes the convergence rate.

\section{Optimal polynomial filters}
\label{sec:polopt}

\subsection{Polynomial filtering for static network topologies}\label{sec:pffixedtopology}

We are now interested in finding the polynomial that
leads to the fastest convergence of linear iteration described in eq.
(\ref{eq:distliniter2}), for a given weight matrix
$W$ and a certain degree $k$. For notational ease, we call $W_p =
p_k(W)= \sum_{i=0}^k \alpha_i W^i$. According to Theorem
\ref{thm:convfixed}, the optimal polynomial is the one that
minimizes the second largest eigenvalue of $W_p$, i.e., $\rho
\big( W_p - \frac{\textbf{1} \textbf{1}^\top}{n} \big)$.
Therefore, we need to solve an optimization problem where the
optimization variables are the $k+1$ polynomial coefficients
$\alpha_0,\ldots,\alpha_k$ and the objective function is the
spectral radius of $W_p - \frac{\textbf{1} \textbf{1}^\top}{n}$.

\begin{center}
\fbox{\makebox{
\begin{tabular}{l}
Optimization problem: \textbf{OPT1} \\
$ \min_{\alpha \in \mathbb{R}^{k+1}} ~\rho
\big(\sum_{i=0}^k \alpha_i W^i - \frac{\textbf{1} \textbf{1}^\top}{n}  \big)$\\
subject to \\
\quad $\big( \sum_{i=0}^k \alpha_i W^i \big) \textbf{1} = \textbf{1}$.
\end{tabular}
}}
\end{center}

Interestingly, the optimization problem OPT1 is convex. First, its objective function is convex, as stated in Lemma
\ref{lemma:objfunconvexity}.
\begin{lemma}\label{lemma:objfunconvexity}
For a given symmetric weight matrix $W$ and degree $k$, $\rho \big(\sum_{i=0}^k \alpha_i W^i -
\frac{\textbf{1} \textbf{1}^\top}{n} \big)$ is a convex function of the polynomial
coefficients $\alpha_i$'s.
\end{lemma}
\textit{Proof}: Let $\beta, \gamma \in \mathbb{R}^{k+1}$ and $0
\leq \theta \leq 1$. Since $W$ is symmetric, $\sum_{i=0}^k
\alpha_i W^i$ is also symmetric. Hence, the spectral radius is
equal to the matrix 2-norm. Thus, we have
\begin{eqnarray*}
&& \rho \big( \sum_{i=0}^k (\theta \beta_i + (1-\theta)\gamma_i) W^i- \frac{\textbf{1} \textbf{1}^\top}{n} \big)  \\
&=& \Big\| \sum_{i=0}^k (\theta \beta_i + (1-\theta)\gamma_i) W^i - \frac{\textbf{1} \textbf{1}^\top}{n} \Big\|_2 \\
&=& \Big\| \theta \Big( \sum_{i=0}^k \beta_i  W^i - \frac{\textbf{1} \textbf{1}^\top}{n} \Big) + (1-\theta) \Big( \sum_{i=0}^k \gamma_i W^i -\frac{\textbf{1} \textbf{1}^\top}{n} \Big) \Big\|_2 \\
&\leq& \theta \Big\|  \sum_{i=0}^k \beta_i  W^i - \frac{\textbf{1} \textbf{1}^\top}{n} \Big\|_2 + (1-\theta) \Big\| \sum_{i=0}^k \gamma_i W^i -\frac{\textbf{1} \textbf{1}^\top}{n} \Big\|_2 \\
&\leq& \theta \rho \Big( \sum_{i=0}^k \beta_i  W^i - \frac{\textbf{1} \textbf{1}^\top}{n} \Big) + (1-\theta) \rho \Big( \sum_{i=0}^k \gamma_i W^i -\frac{\textbf{1} \textbf{1}^\top}{n} \Big),
\end{eqnarray*}
which proves the Lemma.
In addition, the constraint of OPT1 is linear which implies that the set of feasible $\alpha_i$'s is
convex. As OPT1 minimizes a convex function over a convex set, the optimization problem is indeed convex.

In order to solve OPT1, we use an auxiliary variable $s$ to bound
the objective function, and then we express the spectral radius
constraint as a linear matrix inequality (LMI). Thus, we need to
solve the following optimization problem.

\begin{center}
\fbox{\makebox{
\begin{tabular}{l}
Optimization problem: \textbf{OPT2} \\
$\min_{s \in \mathbb{R}, \alpha \in \mathbb{R}^{k+1}} ~s$\\
subject to \\
\quad $-sI \preceq \sum_{i=0}^k \alpha_i W^i - \frac{\textbf{1}\textbf{1}^\top}{n}  \preceq sI$, \\
\quad $\big( \sum_{i=0}^k \alpha_i W^i \big) \textbf{1} = \textbf{1}$.
\end{tabular}
}}
\end{center}
Recall that since $W$ is symmetric, $W_p = \sum_{i=0}^k \alpha_i
W^i$ will be symmetric as well. Hence, the constraint $W_p
\textbf{1} = \textbf{1}$ is sufficient to ensure that $\textbf{1}$
will be also a left eigenvector of $W_p$. The spectral radius
constraint, \[-sI \preceq W_p -
\frac{\textbf{1}\textbf{1}^\top}{n}  \preceq sI\] ensures that
that all the eigenvalues of $W_p$ other than the first one, are
less or equal to $s$. Due to the LMI, the above optimization
problem becomes equivalent to a semi-definite program (SDP) \cite{BoydVan-book}. SDPs
are convex problems and can be globally and efficiently solved. The solution to OPT2 is therefore computed efficiently in practice, where the SDP only has a moderate number of $k+2$ unknowns (including $s$).

\subsection{Polynomial filtering for dynamic network
topologies}\label{sec:pfrandomtopology}

We extend now the idea of polynomial filtering to dynamic network topologies. Theorem \ref{thm:convdyn} suggests that the
convergence rate in the random network topology case is governed
by $E \Big[ \rho \big(W - \frac{\textbf{1} \textbf{1}^\top}{n}
\big) \Big]$. Since $W$ depends on a dynamic edge set, $p_k(W)$ now becomes stochastic.
Following the same intuition as above, we could form an
optimization problem, similar to OPT1, whose objective function
would be $E[\rho\big(\sum_{i=0}^k \alpha_i W^i - \frac{\textbf{1}
\textbf{1}^\top}{n}  \big)]$.
Although this objective function can be shown to be convex, its evaluation is hard and typically requires several Monte
Carlo simulations steps.

Recall that the convergence rate of eq. (\ref{eq:distliniter2dyn}) is related to the second
largest eigenvalue of $E[W]$, which is much easier to
evaluate. Let $\overline{W}$ denote the average weight matrix $E[W]$. We then observe that
\[
E\Big[\rho\big(W - \frac{\textbf{1} \textbf{1}^\top}{n} \big)
\Big] \geq \rho \Big(\overline{W} - \frac{\textbf{1}
\textbf{1}^\top}{n} \Big),
\]
which is due to Lemma \ref{lemma:objfunconvexity} and Jensen's
inequality. The above inequality implies that the spectral radius $\rho \Big(\overline{W} -
\frac{\textbf{1} \textbf{1}^\top}{n} \Big)$ shall be small in order
$E\Big[\rho\big(W - \frac{\textbf{1} \textbf{1}^\top}{n} \big)
\Big]$ to be small too. Additionally, the authors  provide experimental evidence in
\cite{KarMoura.TSP.07}, which indicates that $\rho
\Big(\overline{W} - \frac{\textbf{1} \textbf{1}^\top}{n} \Big)$
seems to be closely related to the convergence rate of eq.
(\ref{eq:distliniter2dyn}).

Based on the above facts, we propose to build our polynomial
filter based on $\overline{W}$. Hence, we formulate the following
optimization problem for computing the polynomial coefficients
$\alpha_i$'s in the random network topology case.

\begin{center}
\fbox{\makebox{
\begin{tabular}{l}
Optimization problem: \textbf{OPT3} \\
$ \min_{\alpha \in \mathbb{R}^{k+1}} ~\rho
\big(\sum_{i=0}^k \alpha_i \overline{W}^i - \frac{\textbf{1} \textbf{1}^\top}{n}  \big)$\\
subject to \\
\quad $\big( \sum_{i=0}^k \alpha_i \overline{W}^i \big) \textbf{1}
= \textbf{1}$.
\end{tabular}
}}
\end{center}
OPT3 could be viewed as the analog of OPT1 for the case of dynamic
network topology. The main difference is that we work on
$\overline{W}$, whose eigenvalues can be easily obtained. Using again the auxiliary variable $s$, we
reach the following formulation for obtaining the $\alpha_i$'s.

\begin{center}
\fbox{\makebox{
\begin{tabular}{l}
Optimization problem: \textbf{OPT4} \\
$\min_{s \in \mathbb{R}, \alpha \in \mathbb{R}^{k+1}} ~s$\\
subject to \\
\quad $-sI \preceq \sum_{i=0}^k \alpha_i \overline{W}^i - \frac{\textbf{1}\textbf{1}^\top}{n}  \preceq sI$, \\
\quad $\big( \sum_{i=0}^k \alpha_i \overline{W}^i \big) \textbf{1}
= \textbf{1}$.
\end{tabular}
}}
\end{center}

Once $\overline{W}$ has been computed, this optimization problem is solved efficiently by a SDP similarly to the case of static networks.

\section{Simulation Results}
\label{sec:experiments}

\subsection{Setup}

In this section, we provide simulation results which show the
effectiveness of the polynomial filtering methodology. First we introduce a few
weight matrices that have been extensively used in the distributed
averaging literature. Suppose that $d(i)$ denotes the degree of
the $i$-th sensor. It has been shown in
\cite{XiaoBoyd.04,XiBoLa.05} that iterating eq.
(\ref{eq:distliniter2}) with the following matrices leads to
convergence to $\mu \textbf{1}$.
\begin{itemize}
\item \textit{Maximum-degree} weights. The Maximum-degree weight
matrix is
\begin{equation}\label{eq:maxdeg}
W_{ij} =
\begin{cases}
\frac{1}{n}         & \textrm{if} ~(i,j) \in \mathcal{E}, \\
1-\frac{d(i)}{n}  & i=j, \\
0                   & \textrm{otherwise}.
\end{cases}
\end{equation}
\item \textit{Metropolis} weights. The Metropolis weight matrix is
\begin{equation}\label{eq:metropolis}
W_{ij} =
\begin{cases}
\frac{1}{1+\max \{d(i),d(j)\}}         & \textrm{if} ~(i,j) \in \mathcal{E}, \\
1- \sum_{(i,k) \in \mathcal{E}} W_{ik} & i=j, \\
0                   & \textrm{otherwise}.
\end{cases}
\end{equation}
\item \textit{Laplacian} weights. Suppose that $A$ is the
adjacency matrix of $\mathcal{G}$ and $D$ is a diagonal matrix
which holds the vertex degrees. The Laplacian
matrix is defined as $L = D-A$ and the Laplacian weight matrix is
defined as
\begin{equation}\label{eq:Lapweight}
W = I - \gamma L,
\end{equation}
where the scalar $\gamma$ must satisfy $\gamma < 1/d_{\max}$
\cite{XiaoBoyd.04}.
\end{itemize}

The sensor networks are built using the random geographic graph model
\cite{GuptaKumar.00}. In particular, we place $n$ nodes uniformly
distributed on the 2-dimensional unit area. Two nodes are adjacent
if their Euclidean distance is smaller than $r =\sqrt{\frac{\log(N)}{N}}$ in order to
guarantee connectedness of the graph with high probability
\cite{GuptaKumar.00}.

Finally, the SDP programs for optimizing the polynomial filters are solved in Matlab using the SeDuMi \cite{Sturm.02}
solver\footnote{Publically available at:
\texttt{http://sedumi.mcmaster.ca/}}.

\subsection{Static network topologies}

\begin{figure*}[!t]
\begin{center}\mbox{
     \subfigure[SDP polynomial filter $p_k(\lambda), ~\lambda_{\min} \leq \lambda \leq 1$]{\label{fig:polfiltmaxdeg}\includegraphics[width=3in]{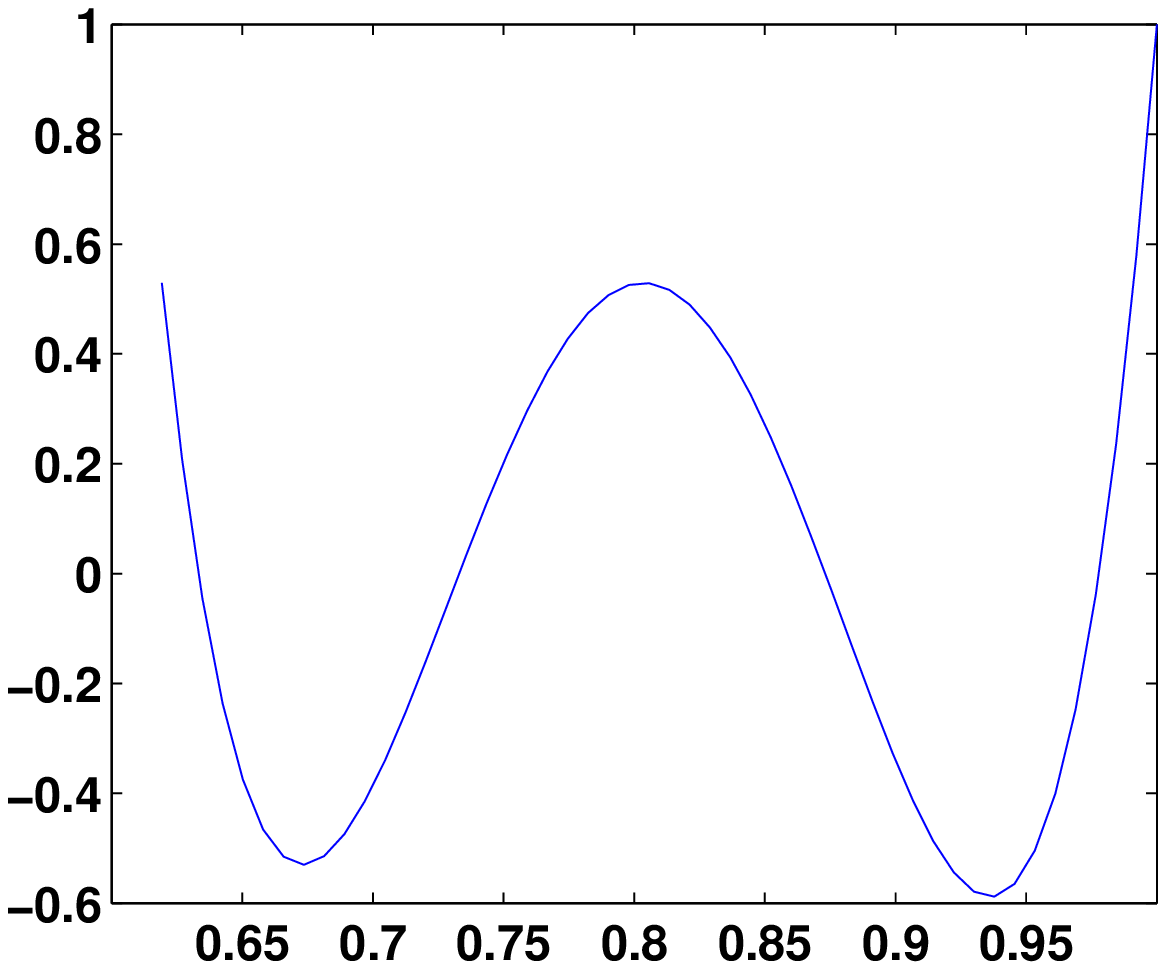}}
     \subfigure[Effect on the spectrum $\lambda(W)$]{\label{fig:eigsmaxdeg}\includegraphics[width=3in]{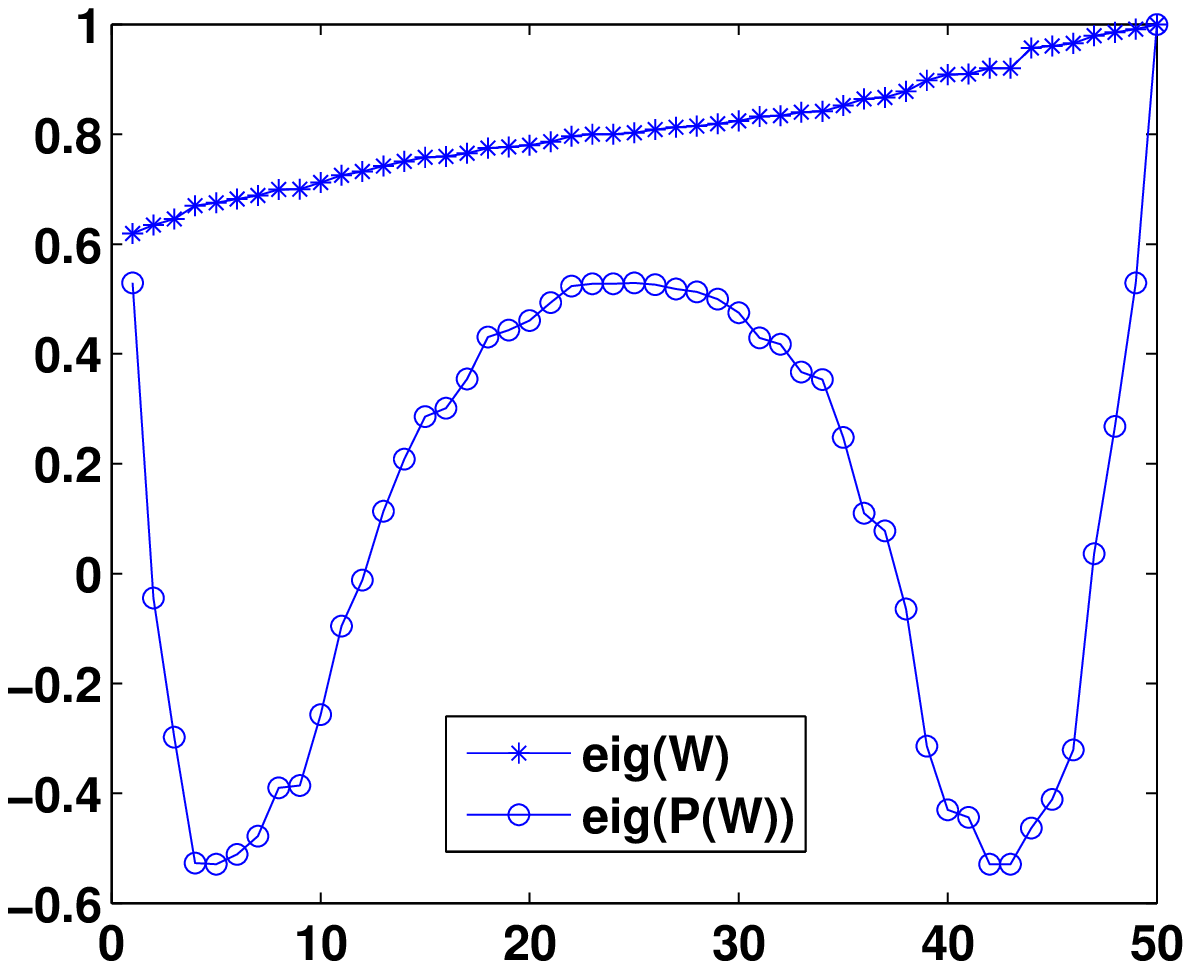}}
     }
\end{center}
     \caption{Effect of polynomial filtering on the spectrum of the Maximum-degree matrix of eq. (\ref{eq:maxdeg}).}
     \label{fig:PFeffect}
\end{figure*}

We illustrate first the effect of polynomial filtering on the spectrum of $W$.
We build a network of $n=50$ sensors and we apply polynomial
filtering on the Maximum-degree weight matrix $W$, given in
(\ref{eq:maxdeg}). We use $k=4$ and we solve the optimization
problem OPT2 using the Maximum-degree matrix $W$ as input. Figure
\ref{fig:polfiltmaxdeg} shows the obtained polynomial filter
$p_k(\lambda)$, when $\lambda \in [\lambda_{\min}(W),1]$. Next, we
apply the polynomial on $W$ and Figure \ref{fig:eigsmaxdeg} shows
the spectrum of $W$ before (star-solid line) and after
(circle-solid line) polynomial filtering, versus the vector index.
Observe that polynomial filtering dramatically increases the
spectral gap $|1 - \lambda_2|$, which further leads to accelerating the distributed consensus, as we show in the
simulations that follow.

\begin{figure*}[!t]
\begin{center}\mbox{
     \subfigure[SDP polynomial filtering]{\label{fig:exper6SDP}\includegraphics[width=3in]{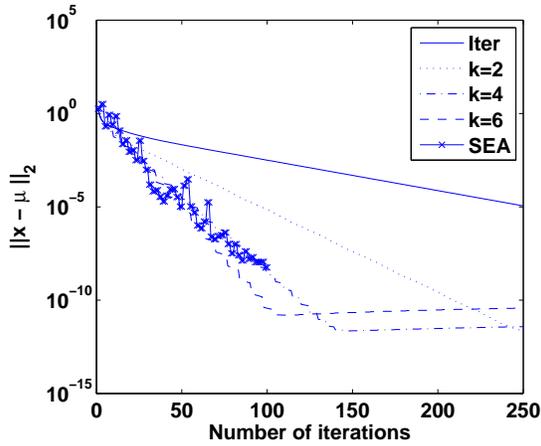}}
     \subfigure[Newton polynomial filtering]{\label{fig:exper6Herm}\includegraphics[width=3in]{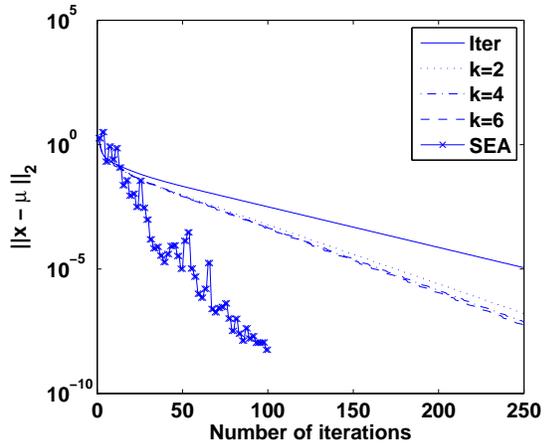}}
     }
\end{center}
     \caption{Behavior of polynomial filtering for variable degree $k$ on fixed topology using the Laplacian weight matrix.}
     \label{fig:fixedtopologyvariablek}
\end{figure*}

\begin{figure*}[!t]
\begin{center}\mbox{
     \subfigure[Max-degree weight matrix]{\label{fig:exper1maxdeg}\includegraphics[width=3in]{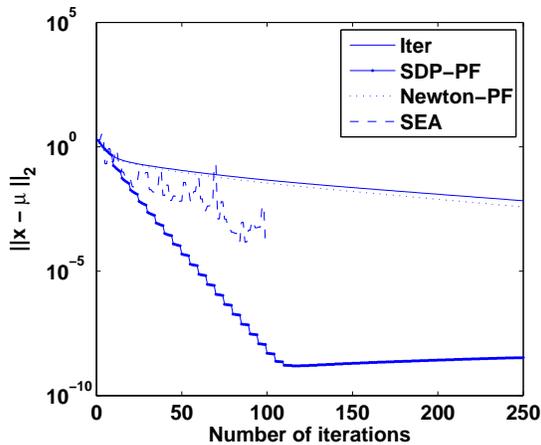}}
     \subfigure[Metropolis weight matrix]{\label{fig:exper1metr}\includegraphics[width=3in]{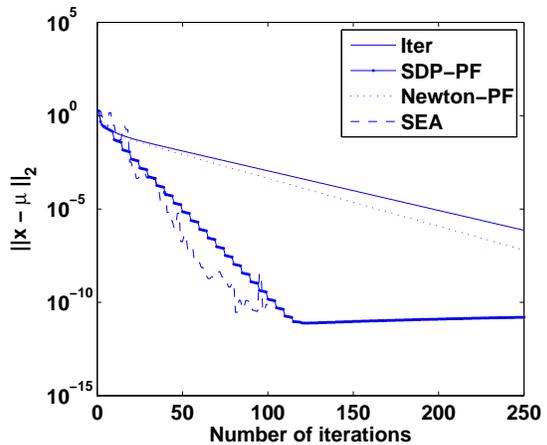}}
     }
\end{center}
     \caption{Comparison between Newton's polynomial and SDP polynomial ($k=4$) on fixed topology.}
     \label{fig:fixedtopology}
\end{figure*}

Then we compare the performance of the different distributed consensus algorithms, with all the aforementioned weight matrices; that is,
Maximum-degree, Metropolis and Laplacian weight matrices for
distributed averaging. We compare both Newton's polynomial and the
SDP polynomial (obtained from the solution of OPT2) with the
standard iterative method, which is based on successive iterations
of eq. (\ref{eq:distliniter2}). For the sake of
completeness, we also provide the results of the scalar epsilon
algorithm (SEA) that uses all previous estimates \cite{KokFro.07.SPL}.

First, we explore the behavior of polynomial filtering methods
under variable degree $k$ from 2 to 6 with step 2. We use the
Laplacian weight matrix for this experiment. Figures
\ref{fig:exper6SDP} and \ref{fig:exper6Herm} illustrate the
evolution of the absolute error $\|x_t - \mu \textbf{1}\|_2$
versus the iteration index $t$, for polynomial filtering with SDP
and Newton's polynomials respectively. We also provide the curve
of the standard iterative method as a baseline. Observe first that
both polynomial filtering methods outperform the standard method
by exhibiting faster convergence rates, across all values of $k$.
Notice also, that the degree $k$ governs the convergence rate,
since larger $k$ implies more effective filtering and therefore
faster convergence. Finally, the stagnation of the convergence process of the SDP polynomial filtering and large values of $k$ is due to the limited accuracy of the SDP solver.

Next, we show the results obtained with the other two weight
matrices on the same sensor network. Figures
\ref{fig:exper1maxdeg} and \ref{fig:exper1metr} show the
convergence behavior of all methods for the Maximum-degree and
Metropolis matrices respectively. In both polynomial filtering
methods we use a representative value of $k$, namely 4. Notice
again that polynomial filtering accelerates the convergence of the
standard iterative method (solid line). As expected, the optimal
polynomial computed with SDP outperforms Newton's polynomial, which is based on intuitive arguments only.

Finally, we can see from Figures
\ref{fig:fixedtopologyvariablek} and \ref{fig:fixedtopology} that
in some cases the convergence rate is comparable for SEA and SDP
polynomial filtering. Note however that the former uses all
previous iterates, in contrast to the latter that uses only the
$k+1$ most recent ones. Hence, the memory requirements are smaller
for polynomial filtering, since they are directly driven by $k$.
This moreover allows more direct control on the convergence rate,
as we have seen in Fig. \ref{fig:fixedtopologyvariablek}. Interestingly, we see
that the convergence process is smoother with polynomial
filtering, which further permits easy extension to dynamic network
topologies.

\subsection{Dynamic network topologies}

\begin{figure*}[!t]
\begin{center}\mbox{
     \subfigure[$k=1$,~$q=1$]{\label{fig:dynamic_0.2}\includegraphics[width=3in]{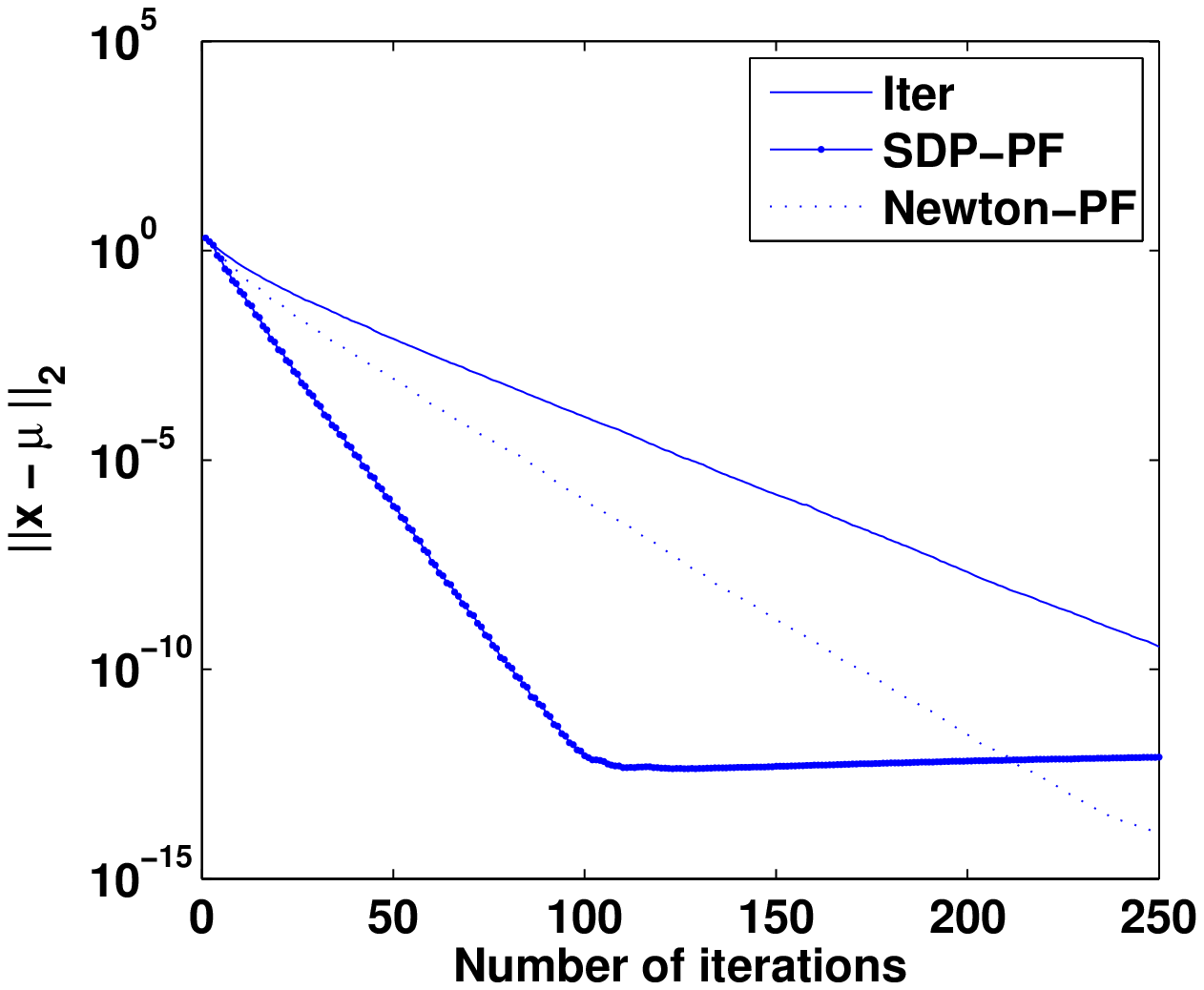}}
     \subfigure[$k=2$,~$q=0.1$]{\label{fig:experiment000}\includegraphics[width=3in]{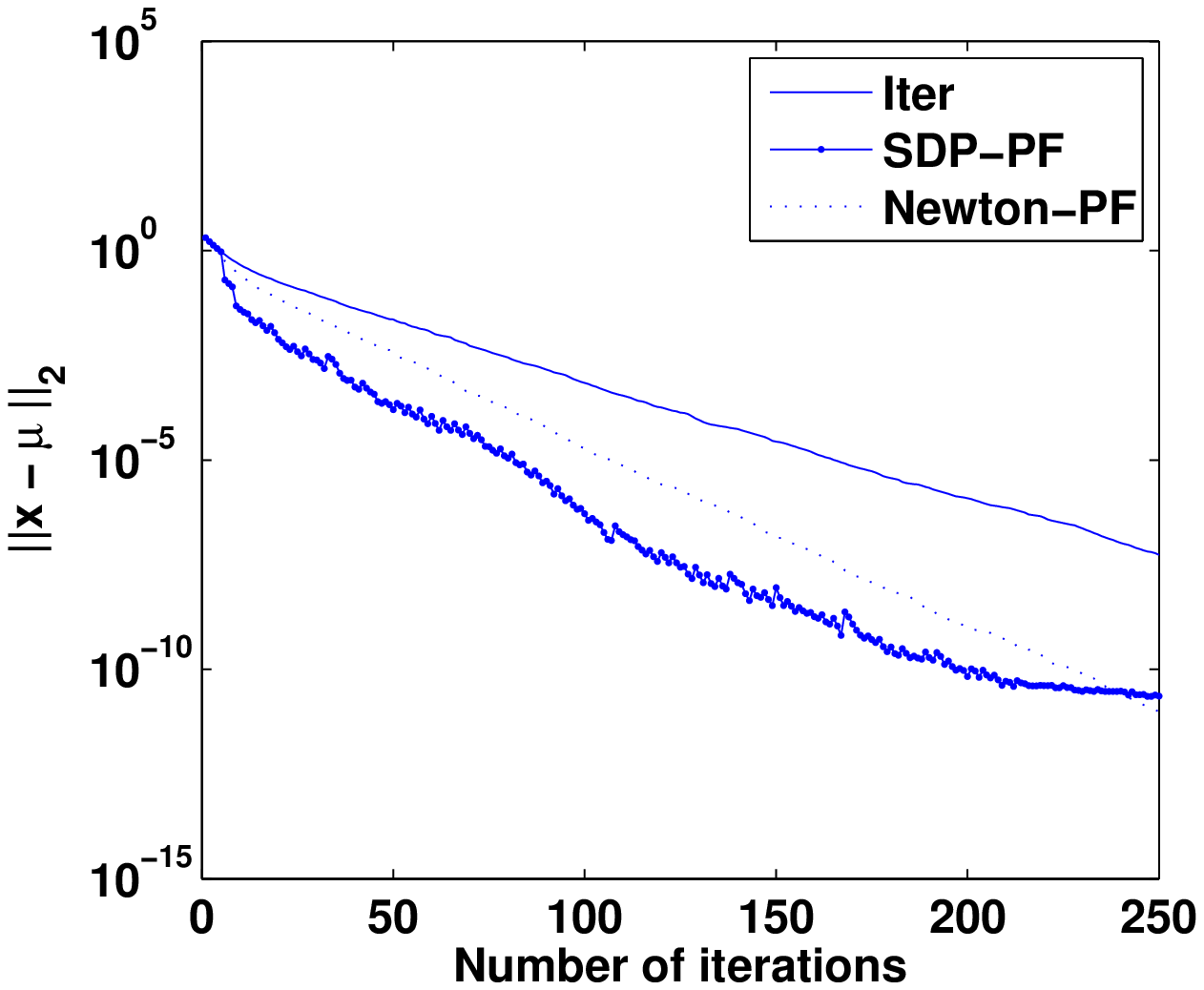}}
     }
     \mbox{
     \subfigure[$k=2$,~$q=0.3$]{\label{fig:experiment000}\includegraphics[width=3in]{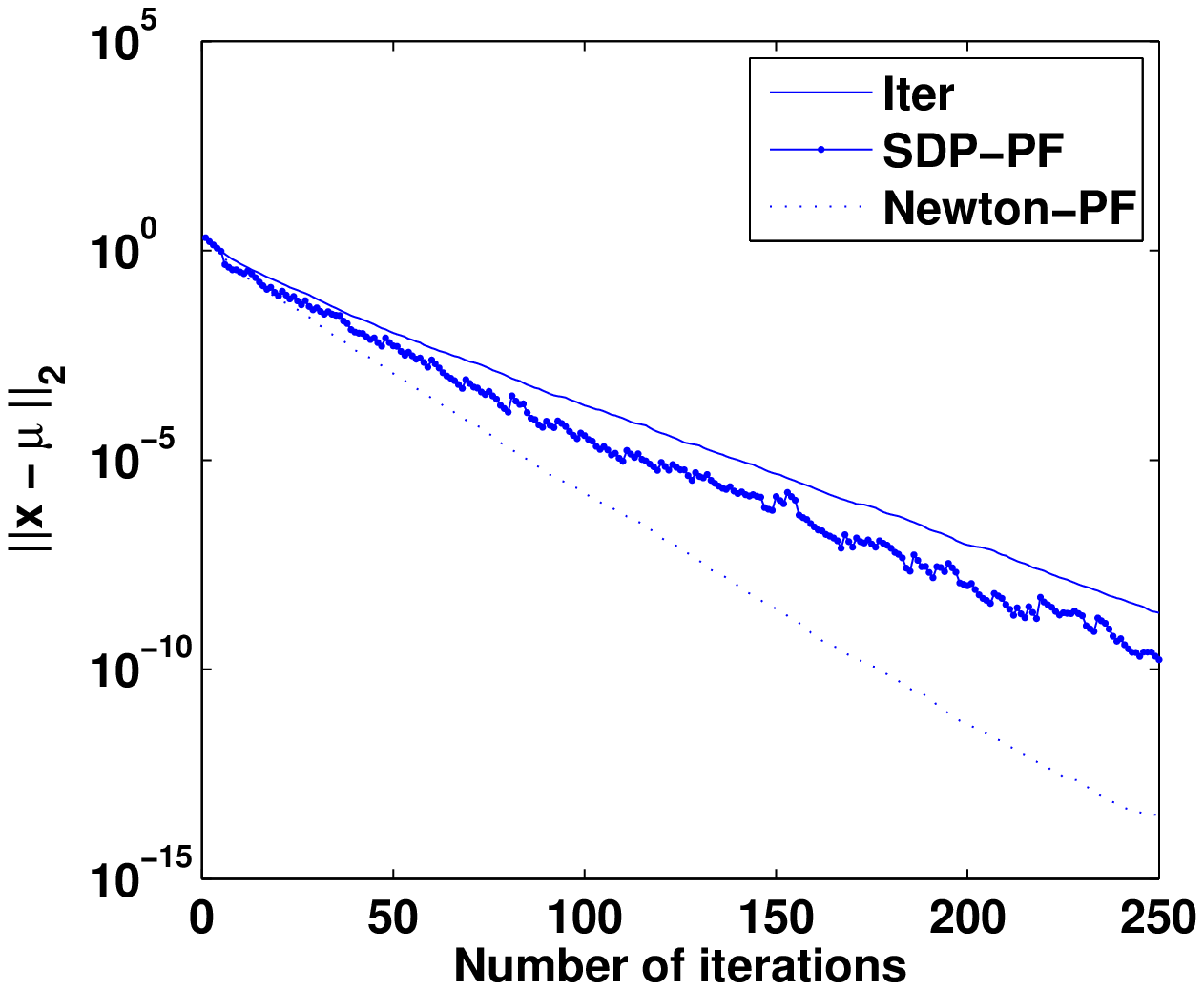}}
     \subfigure[$k=2$,~$q=0.8$]{\label{fig:experiment000}\includegraphics[width=3in]{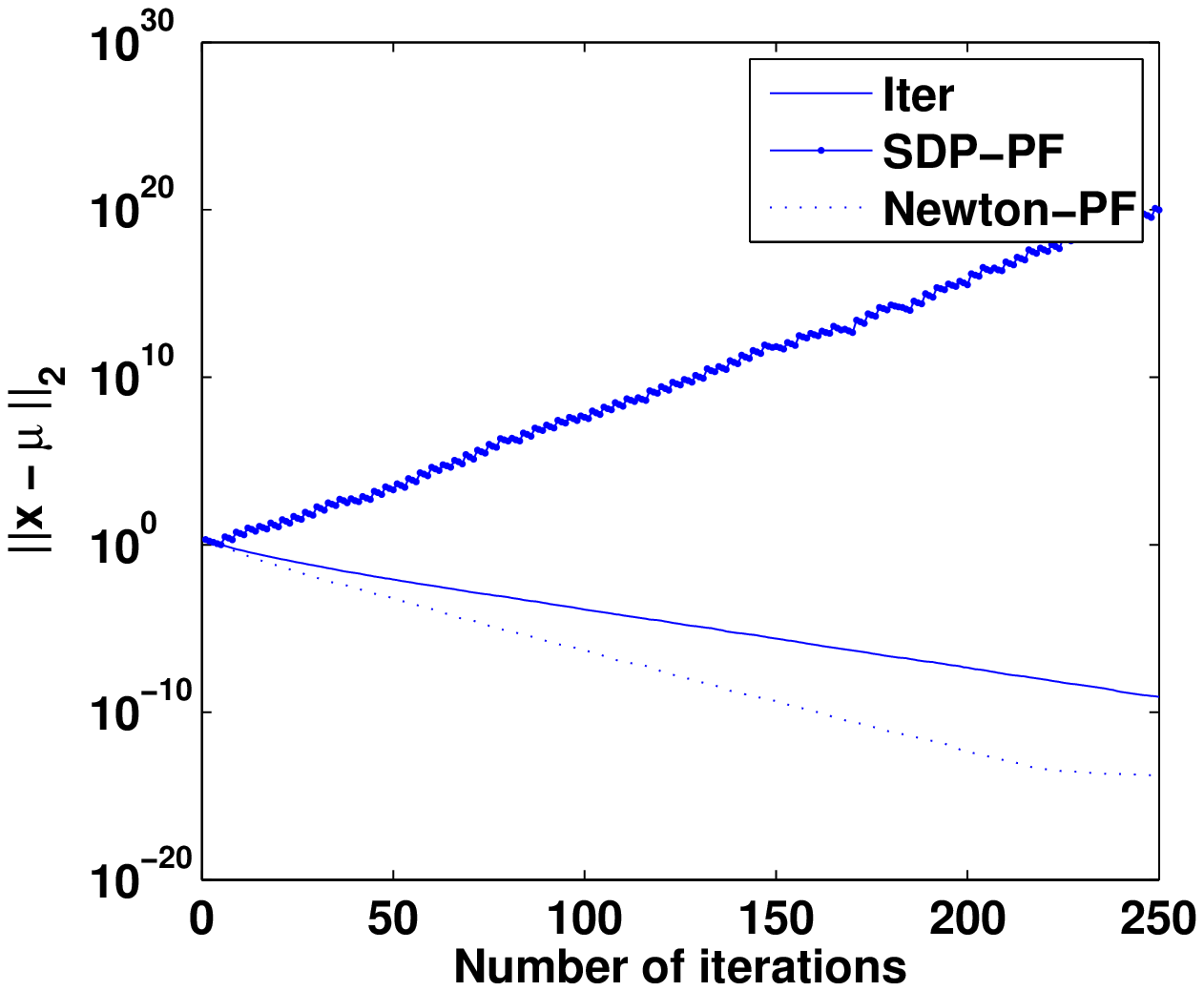}}
     }
\end{center}
     \caption{Random network topology simulations. $q$ denotes the probability that the network changes at each iteration.}
     \label{fig:randomtopology}
\end{figure*}

We study now the performance of polynomial filtering for dynamic networks topologies.
We build a sequence of random networks of $n=50$ sensors, and we assume that in each iteration the network
topology changes independently from the previous iterations, with
probability $q$ and with probability $1-q$ it remains the same as
in the previous iteration. We compare all methods for different values of the probability
$q$. We use the Laplacian weight matrix (\ref{eq:Lapweight}). In
the SDP polynomial filtering method, we solve the SDP program OPT4
(see Sec. \ref{sec:pfrandomtopology}). Fig. \ref{fig:randomtopology}
shows the average performance of polynomial filtering for some
representative values of the degree $k$ and the probability $q$. The average performance is computed using the median
over the 100 experiments. We have not reported the
performance of the SEA algorithm, since it is not robust to
changes of the network topology.

Notice that when $k=1$ (i.e., each sensor uses only its current
value and the right previous one) polynomial filtering accelerates
the convergence over the standard method. At the same time, it
stays robust to network topology changes. Also, observe that in
this case, the SDP polynomial outperforms Newton's polynomial.
However, when $k=2$, the roles between the two polynomial
filtering methods change as the probability $q$ increases. For
instance, when $q=0.8$, the SDP method even diverges. This is
expected if we think that the coefficients of Newton's polynomial
are computed using Hermite interpolation in a given interval and
they do not depend on the specific realization of the underlying
weight matrix. Thus, they are more generic than those of the SDP
polynomial that takes $\overline{W}$ into account, and therefore
less sensitive to the actual topology realization. Algorithms
based on optimal polynomial filtering become inefficient in a
highly dynamic network, whose topology changes very frequently.

\section{Related work}
\label{sec:related}

Several works have studied the convergence rate of distributed
consensus algorithms. In particular, the authors in
\cite{XiaoBoyd.04} and \cite{KarMoura.07,KarMoura.TSP.07} have
shown that the convergence rate depends on the second largest
eigenvalue of the network weight matrix, for fixed and random
networks, respectively. They both use semi-definite programs to
compute the optimal weight matrix, and the optimal topology.

Other works have addressed the consensus problem, and we mention here only the most relevant ones. A. Olshevsky and J. N. Tsitsiklis in \cite{OlshTsi.06} propose two
consensus algorithms for fixed network topologies, which build on
the ``agreement algorithm". The proposed algorithms make use of
spanning trees and the authors bound their worst-case convergence
rate. For dynamic network topologies, they propose an algorithm
which builds on a previously known distributed load balancing
algorithm. In this case, the authors show that the algorithm has a
polynomial bound on the convergence time ($\epsilon$-convergence).

The authors in \cite{HatMes.04} study the convergence properties
of agreement over random networks following the Erd\H{o}s and
R\'{e}nyi random graph model. According to this model, each edge
of the graph exists with probability $p$, independently of other
edges and the value of $p$ is the same for all edges. By
agreement, we consider the case where all nodes of the graph agree
on a particular value. The authors employ results from stochastic
stability in order to establish convergence of agreement over
random networks. Also, it is shown that the rate of convergence is
governed by the expectation of an exponential factor, which
involves the second smallest eigenvalue of the Laplacian of the
graph.

Gossip algorithms have also been applied successfully to solving distributed averaging problems.
In \cite{BoydGhoshPrabShah.06} provide convergence results on randomized gossip algorithm in both
synchronous and asynchronous settings. Based on the obtained
results, they optimize the network topology (edge formation
probabilities) in order to maximize the convergence rate of
randomized gossip. This optimization problem is also formulated as a
semi-definite program (SDP).
In a recent study, the authors in \cite{DimSarJain.08} have been able
to improve the standard gossip protocols in cases where the
sensors know their geometric positions. The main idea is to
exploit geographic routing in order to aggregate values among
random nodes that are far away in the network.

Under the same assumption of knowing the geometric positions of
the sensors, the authors in \cite{LiDai-TIT.07} propose a fast
consensus algorithm for geographic random graphs. In particular,
they utilize location information of the sensors in order to
construct a nonreversible lifted Markov chain that mixes faster
than corresponding reversible chains. The main idea of lifting is
to distinguish the graph nodes from the states of the Markov chain
and to ``split" the states into virtual states that are connected
in such a way that permits faster mixing. The lifted graph is then
``projected" back to the original graph, where the dynamics of the
lifted Markov chain are simulated subject to the original graph
topology. However, the proposed algorithm is not applicable in the
case where the nodes' geographic location is not available.

In \cite{LiDai-ICASSP.07} the authors propose
a cluster-based distributed averaging algorithm, applicable to
both fixed linear iteration and random gossiping. The induced
overlay graph that is constructed by clustering the nodes is
better connected relatively to the original graph; hence, the
random walk on the overlay graph mixes faster than the
corresponding walk on the original graph. Along the same lines, K.
Jung et al. in \cite{JungShah.06}, have used nonreversible lifted
Markov chains to accelerate consensus. They use the lifting scheme
of \cite{ChenLovPak.99} and they propose a deterministic gossip
algorithm based on a set of disjoint maximal matchings, in order
to simulate the dynamics of the lifted Markov chain.

Finally, even if we have mostly considered synchronous algorithms
in this paper, it is worth mentioning that the authors in
\cite{MeSpanosPongLowMurray.07} propose two asynchronous
algorithms for distributed averaging. The first algorithm is based
on blocking (that is, when two nodes update their values they
block until the update has been completed) and the other algorithm
drops the blocking assumption. The authors show the convergence of
both algorithms under very general asynchronous timing
assumptions. Moreover, the authors in \cite{MoallemiVanRoy.06}
propose \textit{consensus propagation}, which is an asynchronous
distributed protocol that is a special case of belief propagation.
In the case of singly-connected graphs (i.e., connected with no
loops), synchronous consensus propagation converges in a number of
iterations that is equal to the diameter of the graph. The authors
provide convergence analysis for regular graphs.

\section{Conclusions}\label{sec:conclusion}

In this paper, we proposed a polynomial filtering methodology in
order to accelerate distributed average consensus in both fixed
and random network topologies. The main idea of polynomial
filtering is to shape the spectrum of the polynomial weight matrix
in order to minimize its second largest eigenvalue and
subsequently increase the convergence rate. We have constructed
semi-definite programs to compute the optimal polynomial
coefficients in both static and dynamic networks. Simulation
results with several common weight matrices have shown that the
convergence rate is much higher than for state-of-the-art
algorithms in most scenarios, except in the specific case of
highly dynamic networks and small memory sensors.

\section{Acknowledgements}
The first author would like to thank prof. Yousef Saad for the
valuable and insightful discussions on polynomial filtering.

\bibliographystyle{IEEEtran}

\end{document}